\documentclass[preprint,showpacs,preprintnumbers,amsmath,amssymb]{revtex4}

\usepackage{graphicx}
\usepackage{dcolumn}
\usepackage{bm}

\begin{document}

\title{Area, ladder symmetry, degeneracy and fluctuations of a horizon}

\author{Mohammad H. Ansari}
\email{mansari@perimeterinstitute.ca}
\affiliation{University of Waterloo, Waterloo, On, Canada N2L 3G1 }%
\affiliation{Perimeter Institute, Waterloo, On, Canada N2L 2Y5}%
\date{\today}
\newcommand{\beq}{\begin{equation}}
\newcommand{\eeq}{\end{equation}}
\newcommand{\barr}{\begin{array}}
\newcommand{\earr}{\end{array}}
\newcommand{\ssz}{\scriptsize}
\newcommand{\amin}{a_{\mathrm{min}}}
\newcommand{\zmin}{\zeta_{\mathrm{min}}}

\begin{abstract}
Loop quantum gravity admits a kind of area quantization that is
characterized by three quantum numbers. We show the complete
spectrum of area is the union of equidistant subsets and a universal
reformulation with fewer parameters is possible. Associated with any
area there is also another number that determines its degeneracy.
One application is that a quantum horizon manifests harmonic modes
in vacuum fluctuations. It is discussed the physical fluctuations of
a space-time horizon should include all the excluded area
eigenvalues, where quantum amplification effect occurs. Due to this
effect the uniformity of transition matrix elements between near
levels could be assumed. Based on these, a modification to the
previous method of analyzing the radiance intensities is presented
that makes the result one step further precise. A few of harmonic
modes appear to be extremely amplified on top of the Hawking's
radiation. They are expected to form a few brightest lines with the
wavelength not larger than the black hole size.

\end{abstract}

\maketitle

\tableofcontents

\section{Introduction}

So far the main consequence of area quantization in loop quantum
gravity has been the removal of classical gravitational
singularities \cite{Bojowald:2001xe} as well as determining the
isolated horizon entropy \cite{Ashtekar:1997yu}. The predicted
generic exit of scale factor from an inflation sector into a
Friedman universe in a loop quantized minisuperspace is at present
in agreement with standard inflation models. This quantum phenomena,
which comes from a quantum correction in the inhomogeneous
Hamiltonian constraint, is through elementary area variable whose
value should be determined by an underlying inhomogeneous state.
Area is an elementary operator in loop quantum gravity because in
the classical limits it is directly related to the densitized triad
as a canonical variable. In this note, we study two previously
unknown properties of area quantization that further clarify the
understanding of this operator. Firstly, the area eigenvalues
possess a symmetry that its spectrum is the union of different
evenly spaced subsets. Secondly, the eigenvalues are substantially
degenerate such that in larger area the degeneracy increases. Due to
the presence of a huge class of completely tangential excitations on
a surface  different regions of the surface are distinguishable.
These together result in degeneracy increasing in a way that with
any eigenvalue a finite exponentially proportional to area
degeneracy is associated. One application is in area fluctuations of
a collapsing star. It is discussed a trustworthy analysis of area
fluctuations in a space-time horizon must include all those excluded
quanta from a quantized isolated horizon \cite{Ashtekar:1997yu}.
Having recognized the quantum amplification effect during
transitions, the density matrix elements can be considered uniform
in near levels. The black hole undergoes a thermal fluctuations and
harmonic modes resonate. Using these properties, a modification
method to the previous analysis of the intensities
\cite{Ansari:2006vg} is introduced that makes the result one step
further precise. The major result is that the fluctuations in the
dominant configuration with minimal quantum of area is mostly
amplified by the black hole such that a few sharp and bright lines
appear on top of Hawking's radiation. These modes cannot be seen in
the wavelength larger than the size of black hole. In summary, by
the use of a few main assumptions from black hole studies, loop
quantum gravity, a non-perturbative background independent approach
to quantum gravity, becomes testable much above the Planck scales if
quantum primordial black holes are ever found.

\section{\label{sec:area}Area}


In this note we choose to define a surface by a coordinate
condition. The quantization of a 3-manifold is obtained by
quantizing the holonomy configuration space on embedded graphs in a
spatial manifold. The sub-graphs whose nodes lie on a surface are
basis for defining the quantum state of the surface. Densitized
conjugate momenta possess full information of the surface metric and
consequently the surface area, \cite{Ashtekar:surface}.

Consider a vertex lying on a surface with total upper side spin
$j_u$, bottom side spin $j_d$, and completely tangential edges of
total spin $j_t$ on the surface. The quantum of area in this state
depends on the upper and lower spins as well as the total tangent
vector induced by these two on the surface, $j_{u+d}$, \beq
\label{eq. A} a = a_o \sqrt{2f(j_u) + 2f(j_d) - f(j_{u+d})}, \eeq
where $f(x)=x(x+1)$, $a_o := 4 \pi \gamma \ell_P^2 $, $\gamma$ the
Barbero-Immirzi parameter, and $j_{u+d} \in [|j_u - j_d|, \ldots,
j_u+j_d-1, j_u+j_d]$. Note that the completely tangential edges do
not contribute in the area.

Consider a closed underlying surface dividing the 3-manifold into
two completely disjoint sectors and not bounded by a boundary. A few
additional vertices are needed in order to close this quantum state.
This introduces two additional constraints on the states, namely: $
\sum_{\alpha} j_u^{(\alpha)}  \in  \mathrm{Z}^+$ and $\sum_{\alpha}
j_d^{(\alpha)}  \in  \mathrm{Z}^+$, where $\alpha$ labels all the
residing vertices on the surface, \cite{Ashtekar:surface}.

\section{Ladder symmetry}


In SO(3) group representation spins are integers. Therefore in
(\ref{eq. A}) the right side can be written as a positive integer
number: $m \doteq f(j_u) + f(j_d) - \frac{1}{2}f(j_{u+d}) =
\frac{1}{2} \left(a/a_o\right)^2$. This number due to the following
proof is in fact \emph{any} natural number. Suppose $j_u \geq j_d$
and the difference of them is a positive integer $n = j_u - j_d$.
Restricting to the subset $\mathrm{M}^*$ of $j_{u+d} = j_u + j_d$,
it is easy to verify the generator of this subset is $n(n+1)/2 +
j_{d}$. The first term, a triangular number, is a positive integer.
The second term is independent of $n$ and in principle takes any
positive integer value. Therefore the set of all $\mathrm{M}^*$
corresponding to the states with $j_d=\{1,2,3,4,\cdots\}$ is
equivalent to natural numbers; $\mathbb{N}  \equiv \{\mathrm{M}^*
\}/ \mathcal{R }$, where $/ \mathcal{R}$ stands for the modulation
of repetition (or in a simple word different copies of one number
are identified). Since $m$ in general is a positive integer, any
other subset fits into $\mathbb{N}$. Consequently, an irreducible
reformulation of area when all copies of numbers are identified is
possible by one quantum number, $a = a_o \sqrt{2 n\ }$, where $n \in
\mathbb{N}$.

The spectrum of area modulo repetitions in SU(2) group
representations is impossible to reformulate by one parameter;
however, it is possible by two in the following form: $a =
\frac{1}{2} a_o \sqrt{\zeta}\ n$, for any discriminant of positive
definite form $\zeta$ and any positive integer $n$,
\cite{Ansari:2006cx}.

A universal reformulation is thus possible if one rewrites the SO(3)
irreducible reformulation as a reducible one by two parameters. In
the followings it is shown that any integer $c$ can be represented
uniquely by $c=\zeta n^2$ where $\zeta$ is a square-free number and
$n \in \mathbb{N}$. A positive integer that has no perfect square
divisors except 1 is called square-free (or quadratfrei) number. In
other words it is a number whose prime decomposition contains no
repeated factors; for instance 15 is square-free but 18 is not. Now
consider an integer $c$ containing $s$ different prime factors $p_1,
p_2, \cdots, p_s$ each repeated $n_1, n_2, \cdots, n_s$ times,
respectively; $c = \prod_{i=1}^{s}(p_i)^{n_i}$. The exponents $n_i$
are all positive integers and are either even or odd numbers.
Consider the case that the exponents are all odd numbers, $n_i :=
2m_i+1$. Therefore $c$ can be written in the form of
$(\prod_{k=1}^{s} p_k).(\prod_{l=1}^{s} (p_{l})^{m_{l}})^2$ which
shows the integer $c$ is a multiplication of a square-free part and
a square part. This could be redone for any integer number and the
result is the same decomposition. Since the prime factorization of
every number is unique, so does its decomposition into square and
square-free numbers. Therefore, in SO(3) group the complete set of
quantum area $\{m\}$, which fits into natural numbers, is the
multiplication of a square-free and a square number. In other words,
the quantum of area can be reformulated into $a = a_o \sqrt{2 \zeta
}\ n$. This makes the universal reformulation of area as a function
of $\zeta$ and $n$, \beq \label{eq. universal A} a_n(\zeta) = a_0
\chi \sqrt{\zeta} \ n \eeq for $\forall\ n \in \mathbb{N}$, where in
SO(3) group $\zeta$ is any square-free number $\{1,2,3,5,\cdots\}$
and $\chi := \sqrt{2}$; and in SU(2) group $\zeta$ is the
`discriminant of any positive definite form' $\{3,4,7,8,11,\cdots\}$
and $\chi := 1/2$. The parameters $\chi$ is `the group
characteristic parameter.' Fixing $\zeta$ a generation of evenly
spaced numbers is picked out, thus the parameter $\zeta$ is the
`generational number.' For the purpose of making the rest of this
note easier to read let us rename the first generational number
whose gap between levels is minimal by $\zmin$ and the minimal area
$\amin$.

Note that the term $\sqrt{\zeta}$ is an irrational number in both
groups and in any generation it is unique. Therefore the sum or
difference of any two quanta $a_{n_1}(\zeta_1)$ and
$a_{n_2}(\zeta_2)$ for $\zeta_1 \neq \zeta_2$ is unique and belongs
to none of generations.

\section{Degeneracy}


The spin network states of a surface under the action of area
operator manifest a substantial degeneracy. Consider an $N$-valent
vertex lying on a surface, some of the edges are contained in the
upper side, some in the lower, and some lie completely tangential on
the surface. Given the total spin of upper and lower sectors by
$j_u$ and $j_d$, respectively, a set of area eigenvalues are
generated from a minimum where $j_{u+d}=j_u+j_d$ to a maximum where
$j_{u+d}=|j_u - j_d|$ from eq. (\ref{eq. A}). Changing $j_u$ and
$j_d$ a different finite subset of area is generated whose elements
may or may not coincide with the elements of the other subset of
area eigenvalues. Associated with any area eigenvalue there appears
unexpectedly a finite number of completely \emph{different}
eigenstates. For instance, these states $|j_u=1, j_d=0,
j_{u+d}=1\rangle$, $|j_u=0, j_d=1, j_{u+d}=1 \rangle$, and $|j_u=1,
j_d=1, j_{u+d}=2 \rangle$ correspond to the area $a=\sqrt{2} a_o$.
Counting these states for every eigenvalue a power law correlation
with the size of area appears such that a larger area possesses a
higher degeneracy. This is studied for both $\mathrm{SU(2)}$ and
$\mathrm{SO(3)}$ gauge groups in \cite{Ansari:2006cx}.

On a classical surface there are a finite number of area cells and a
set of degenerate quantum states could be associated with it.
However, this is essential for a background independent theory to
identify only physical states after reducing the redundant gauge-
and diffeomorphism-transformed ones. Gauge invariance by definition
is satisfied in spin network state, but diffeomorphism invariance
should be checked by its imposing on the states. Consider a surface
containing a large number of the same area cell in different
regions. Each cell is a degenerate eigenvalue of area. However, area
operator does not `see' the completely tangential edges of these
degenerate states. By definitions, the number of completely
tangential edges at each vertex could vary from zero to infinity and
when there are many of these excitations at one vertex they accept a
huge spectrum of spins. These various states make the identical cell
configuration on different regions \emph{distinguishable} under the
measurements of other observable operators.

Note that the area of higher levels can be decomposed precisely into
smaller fractions of the same generation (without any
approximation). For example, $a_n = n a_1 = (n-2) a_1 + a_2 =
\cdots$. As it was explained above, these cells are all completely
distinguishable. Therefore the degeneracy of the area eigenvalue
$a_n$ becomes $\Omega_n = g_n + g_{n-1}g_1 + \cdots + (g_1)^n$.
Obviously the dominant term in the sum belongs to the configuration
with maximum number of the area cell $a_1$. Therefore the total
degeneracy of $a_n(\zeta)$ for $n \gg 1$ is: \beq \label{eq deg}
\Omega_n(\zeta) = g_1(\zeta)^n. \eeq

In the classical limits, the dominant configuration of a large
surface is the one occupying the highest possible level of area from
the `first' generation $\zmin$; i. e. $A \approx n \amin$. This
dominant degeneracy is $g_1(\zmin)^n$ and a kinematic entropy can be
associated with it proportional to the area; $S = A (\ln g_1(\zmin)
/ \amin)$. Depending on the type of time evolution of the surface
this entropy may vanish, decrease, increase or remains unchanged in
the course of time. In other words, a classical surface
characterized by its area at each time slice possesses a finite
entropy-like parameter. Space-time horizons as a class of physical
surfaces possess a non-decreasing entropy. In other words their
kinematical entropy in the course of time, due to the second black
hole thermodynamics law, are \emph{physical} entropy. We will show
in the next section such a horizon carries an entropy whose nature
is the total degeneracy of vacuum fluctuation modes responsible for
the thermal radiation of black hole. However, for the aim of this
note on the study of kinematics of fluctuations we disregard here
the issues of defining the Hamiltonian of a quantum horizon based on
spin foam, which is still an open problem.

\section{Fluctuations of a horizon}


Having known a suitable definition for the information flow other
than expansions of geodesic congruences used in general relativity,
one can certainly define a quantum black hole.  However, there are
different definitions of quantum horizons with different properties,
including causal ones. Event horizon is always a null surface by
definition, thus it must satisfy one-way information transfer,
\cite{Beckman:2001qs}. However an event horizon is not locally
defined at all, not even in time. To define it classically, we need
the information of the whole manifold. In canonical quantum gravity,
we need a definition by which we can look at a place in space and
say those photons reaching to us from there must come from a spatial
slice that intersects a space-time horizon. Such local definitions
are in fact those of apparent, trapping, and dynamical horizons,
\cite{Ashtekar:1997yu}. On the other hand, the space-time horizons
are \emph{not} necessarily null. They would be so if we have vacuum
and absence of gravitational radiation. Vacuum can easily be
achieved for a spin network case, but we cannot prevent the local
gravitational degrees of freedom to be excited in the neighborhood
of a space-time horizon. With these gravitational radiation across
the horizon and with positive energy conditions (or vacuum) the
horizon will be \emph{space-like} rather than null. Moreover, the
energy conditions in quantum gravity could not be taken for granted,
even for semiclassical states, as long as violations occur on small
length scales, \footnote{ There are also examples in quantum field
theory on a curved background for how energy conditions can be
violated locally.}. Thus, quantum space-time horizons can become
even \emph{timelike} with a two-way information transfer. As a
consequence, one \emph{cannot} restrict the quantum fluctuations of
horizon area to the subset that is considered in the trapping-based
theories of horizon because the basic assumption underneath those
theories is that a quantum horizon is the extention of a classical
null boundary of space-time in a quantum theory,
\cite{Ashtekar:1997yu}. Physical fluctuations of space-time
horizons, in fact, occurs in a wider spectrum that includes all
excluded quanta of area.

Note that in the Hawking's conception of a black hole radiation,
those modes created in vacuum at past null infinity pass through the
center of a collapsing star, hover around it and come out of it at
future infinity. The outgoing quanta get a thermal statistics from
this incipient (about-to-be-formed) black hole. Quantum fluctuations
of the horizon change this simple picture because the Hawking quanta
will not be able to hover at a nearly fixed distance from the
fluctuating horizon. Bekenstein and Mukhanov postulated an
equidistant spectrum for the horizon area fluctuations in
\cite{Bekenstein:1995ju} and showed concentrating of radiance modes
in discrete lines. In loop quantum gravity as a fundamental
candidate theory of quantum gravity, quantum of area is different
and here its emissive pattern is work out.

During the latest stages of gravitational collapse of a neutral
non-rotating spherical star, all radiatable multipole perturbations
in the gravitational fields are radiated away such that its
classical physics is described only by its horizon area. The energy
associated with this object depends on the area by the relation $ A
= \frac{16 \pi G^2}{c^4} M^2$. The energy fluctuations of a large
space-time horizon are easy to find $\delta M = \gamma \chi \frac{
M_{Pl}}{ 8 M} \sqrt{\zeta} \delta n$. Ladder symmetry classifies the
transitions between levels into: 1) `generational transitions',
those with both initial and final levels belonging to the same
generation, or 2) `inter-generational transitions', with initial and
final levels belonging to two different generations. The
generational transitions produce `harmonic' frequencies proportional
to a fundamental frequency by an integer. Inter-generational
transitions produce `non-harmonic modes'.

In generational transitions, the fundamental frequency is the jump
between two consecutive levels with frequency $\varpi (\zeta) =
(\gamma \chi \sqrt{\zeta})\ \omega_o $, where
$\omega_o:=\frac{c^3}{8GM}$ is the so-called `frequency scale'. For
instance, a black hole of mass $10^{-18} M_{\odot}$ has a horizon of
area about $10^{-29} \mathrm{m}^2$ and a temperature about $10^{11}$
K. The frequency scale is thus of the order of $\sim 10$ keV. Such a
typical hole has a horizon 40 order of magnitude larger than the
Planck length area. Therefore from each harmonic mode there are many
\emph{copies} emitted in the different levels; or in other words
these modes are amplified. On the other hand, since the difference
of two levels of different generations is a unique number, there
exists only \emph{one} copy from each non-harmonic mode in all
possible transitions. This \emph{quantum amplification effect} makes
a black hole condensate its particles production mostly on harmonic
modes. One important consequence is the density matrix elements of
non-harmonic modes can be regarded negligible and therefore the
generational transitions matrix elements can be assumed to be
uniform.

In a transition down the level of a generation, there are two weight
factors: the transition and the population weights. Assume a hole of
large area $A$. When the hole jumps $f$ steps down the ladder of
levels in the generation $\zeta$, it emits a quanta of the frequency
$f\varpi(\zeta)$. This much of radiance energy could also be emitted
in the dominant configuration by radiating
$f\frac{a_1(\zeta)}{\amin}$ quanta of the fundamental frequency
$\varpi(\zmin)$. These two transitions, although are of the same
radiance frequency, appear with different possibilities. The
degeneracy ratio of these two is $\Omega(f
\varpi(\zeta))/\Omega(f\frac{a_1(\zeta)}{\amin} \varpi(\zmin))$ that
gives rise to the definition of `transition weight' $\theta (\zeta,
f) = g_1(\zeta)^{f} g_1(\zmin)^{-f a_1(\zeta) / \amin}$. The second
weight is the population one that comes from a different root. Due
to quantum amplification effect, from each harmonic frequency there
produced many copies in different levels on the generation. This
weight is in fact the number of possible quanta emitting from
different levels with the same frequency. It is easy to verify this
number is $N_{\varpi(\zeta)}-f+1 $ where $N_{\varpi(\zeta)}$ is the
number of copies from the fundamental frequency, and for near level
modes ($ f \ll N_{\varpi(\zeta)}$) it is $\frac{A}{a_1(\zeta)}$. We
absorb constants in normalization factors and the population weight
in near levels becomes $\rho(\zeta):= 1/\sqrt{\zeta}$.

Finally notice that within one generation when a space-time hole
jumps $f$ steps down the ladder of levels, the degeneracy decreases
by a factor of $g_1(\zeta)^{f}$. Having defined the transition and
the population weights, the conditional probability of
$\omega_f(\zeta)$ emission after using (\ref{eq. universal A})
becomes $P(\omega_f(\zeta) | 1 ) = C^{-1} \rho(\zeta) g_1(\zmin)^{-f
\sqrt{\zeta / \zmin}}$, where $C$ is the normalization factor,
\footnote{From normalization $C = \sum_{\zeta} \rho(\zeta)/|1-
g_1(\zmin)^{-\sqrt{\zeta/\zmin}}|$.}.

One can consider a successive emissions and associates a probability
to it as the multiplication of the probability of each emission. The
conditional probability of a $j$ dimensional sequence of different
frequencies becomes $ \prod_{i=1}^j P(\omega_{f_i}(\zeta_i) | 1 )$.
The probability of the sequences to include $k$ emissions out of $j$
to be of the frequency $\omega_{f^*}(\zeta^*)$ (in no matter what
order) while the rest of accompanying emissions are of any value
except this frequency, is $P(k, \omega_{f^*}(\zeta^*); \{
\omega_{f_1}(\zeta_1), \cdots \}|j) = (^j_k) \left[
P(\omega_{f^*}(\zeta^*) | 1 )\right]^k\ $ $\times \prod_{i=1; \
\zeta \neq \zeta^*}^{j-k}
 P(\omega_{f_i}(\zeta_i) | 1 )$. The accompanying
 modes are allowed to accept any frequency except
 $\omega_{f^*}(\zeta^*)$ and therefore the probabilities of any accompanying
frequency should sum. From the definition of $C$, it is easy to find
out in each sum over accompanying modes instead of $\sum_{\omega
\neq \omega^*} P(\omega_{f_i}(\zeta_i) | 1 )$ we can replace $C -
P(\omega_{f^*}(\zeta^*) | 1 )$ that simplifies the probability to
$P(k, \omega_{f^*}(\zeta^*)|j) =
 (_{j}^{k})
 \left[ P(\omega_{f^*}(\zeta^*) | 1 )\right]^k$ $\times
\left[C-P(\omega_{f^*}(\zeta^*) | 1 )\right]^{j-k}$.

Note that a black hole radiates in a `time' sequential order,
\cite{gerlach}. The probabilities of zero and one jump (of no matter
what frequency) in the time interval $\Delta t$ are $P_{\Delta
t}(0)$ and $P_{\Delta t}(1)$, respectively. In the time interval $2
\Delta t$, the probabilities of zero, one, and two jumps are
$P_{\Delta t}(0)^2$, $2 P_{\Delta t}(0) P_{\Delta t}(1)$, and $2
P_{\Delta t}(0) P_{\Delta t} (2) + P_{\Delta t}(1)^2$, respectively.
By induction this is found for higher number of jumps in an interval
and for longer time. A general solution for the equations the
probability of $j$ time-ordered decays in an interval of time
$\Delta t$ is $P_{\Delta t}(j) = \frac{1}{j!} (\frac{\Delta
t}{\tau})^j \exp(\frac{\Delta t}{\tau})$. Multiplying this
probability with $P(k, \omega_{f^*}(\zeta^*)|j)$ and then summing
over all sequence dimensions $j \geq k$, it is easy to manipulate
the total probability of $k$
 emissions with frequency $\omega_{f^*}(\zeta^*)$ to be $P_{\Delta t}(k, \omega_{f^*}(\zeta^*)) = \frac{1}{k!} (x^*_f)^k
 \exp(-x^*_f)$, where $x^*_f = \frac{\Delta t}{C \tau} \rho(\zeta^*) \
g_1(\zmin)^{-f \sqrt{\zeta / \zmin}}$. This indicates the
distribution of the number of quanta emitted in harmonic modes is
Poisson-like.

Let us now look at the distribution of the number of quanta emitted
from a black body radiation. The probability of one emission of
frequency $\omega^*$ is Boltzmann-like; $\pi_{\omega^*} = B
\exp(-\frac{\hbar \omega^*}{kT})$ where $B$ is normalization factor
$B = \sum_{\omega} \pi_{\omega}$. Successive emissions occurs
independently and therefore the probability of a $j$ dimensional
sequence in which $p$ emissions are of the frequency $\omega^*$ is
$\left(^j_p\right) (\pi_{\omega^*})^k \prod_{i}^{j-p} \sum_{\omega_i
\neq \omega*} \pi_{\omega_i}$. The last summation term can be
replaced from the normalization relation by $B-\pi_{\omega^*}$. This
makes the probability equivalent with the black hole emission
probability $P(k, \omega_{f^*}(\zeta^*)|j)$ when
$g_1(\zmin)^{-f^*\sqrt{\zeta^*/\zmin}}$ (i.e. $\exp(-S)$) is
replaced with $\exp(-\frac{\hbar \omega^*}{kT})$. The analogy
indicates that the hole radiation is characterized by Planck's black
body radiation and the temperature matches the black hole
temperature when the Barbero-immirzi parameter is properly defined
for getting the Bekenstein-Hawking entropy. In fact the black hole
is hot and the thermal character of the radiation is entirely due to
the degeneracy of the levels, the same degeneracy (\ref{eq deg})
that becomes manifest as black hole entropy.

 By definition, the intensity of a mode is the total energy
 emitted in that frequency per unit time and area. The average number of
emissive quanta at a typical harmonic frequency is $\overline{k} =
\sum_{k=1}^{\infty} k P_{\Delta t}(k | \omega_{f}(\zeta)) $.
Calculating this summation gives rise to the intensity
\begin{equation}\label{eq. I}
  \frac{I(\omega_{f}(\zeta))}{I_o} =  fg_1(\zmin)^{-f \sqrt{\zeta / \zmin}}
\end{equation}
where $I_o$ is constant.

To estimate the width of lines, we need to compare the average loss
of collapsing star mass in late times with a black body. The average
of time elapsing between two decays is $\bar{t}=\int dt\ t
P_{t}(1)=2\tau$ and its uncertainty is $(\Delta t)^2 = \int dt
(t-\bar{t})^2P_t(1) =3\tau^2$. The average frequency emitted from a
black hole can be shown to be $\bar{\omega} = \omega_o \gamma \chi$,
\footnote{By definition $ \bar{\omega} = \sum_\zeta \sum_f \omega_f
P_{\Delta t}(\{\omega_f | 1\})$.  After using $\sum_n
nx^{n}=x/(1-x)^2$ for $x<1$ and approximating the sum by an integral
with a high upper bound on $\zeta$, the integral gives the same
result when the sum in the definition of $C$ is approximated sum by
integral.} Moreover, the mean value of the number of jumps in
$\Delta t$ is $\bar{j} = \sum_j j P_{\Delta t}(j)$, which becomes
$\frac{\Delta t}{\tau}$. As a consequence, a black hole losses the
ratio of mass $\frac{\Delta \bar{M}}{\Delta t} = -\frac{\hbar
\omega_o \gamma \chi}{c^2 \tau}$ on average. On the other hand, the
nature of a black hole radiation is the same as a black body where
the loss of mean energy is described by Stephan-Boltzman law,
$\frac{\Delta \bar{M}}{\Delta t} = -\frac{\hbar c^4}{15360 \pi G^2
M^2}$. Comparing these two, one finds $\tau = \frac{1920 \pi \gamma
\chi}{ \omega_o}$. According to the uncertainty principle $\Delta E
\Delta t \sim \hbar$, the frequency uncertainty becomes of the order
of a thousandth of the frequency scale $\omega_o$. This shows that
the spectrum lines are indeed very narrow and the various black hole
lines of one generation are unlikely to overlap.

\begin{figure}
\includegraphics[width=8cm]{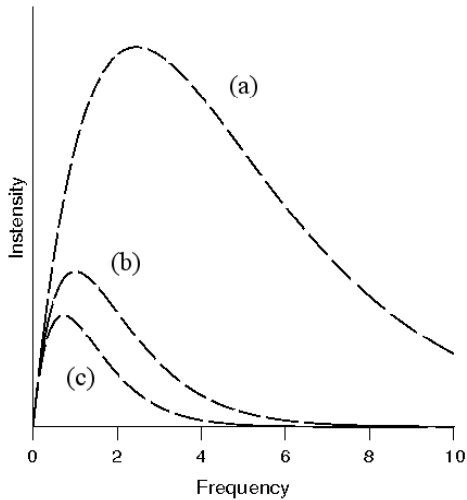}
\caption{The intensity envelope of some
generations.}\label{fig2}\end{figure}

The intensity envelope of the first three generations is plotted in
Fig. (\ref{fig2}), where the envelope (a), (b) and (c) belongs to
the intensity of harmonics in the first, second and third
generations, respectively. It becomes clear that in a generation
with the least gap between levels, the strongest harmonic modes are
amplified. The brightest lines belong to a few of the first
harmonics of the generation $\zmin$. Other than these lines, the
intensity of the rest of harmonics in other generations are
suppressed exponentially. We expect in a low energy spectroscopy a
clear observation of only a few narrow and unblended lines highly on
top of other harmonics. Also we expect these brightest lines appear
in the wavelength not larger than the size of black hole $M$ in
Planck units.

In summary: we showed the quantum of area are substantially
degenerate. The complete spectrum is possible to reformulate into a
universal form with two parameters and more importantly it is the
union of exactly equidistant subsets. The spectrum of radiation due
to these new properties reveals a clear discretization on a few
brightest lines which cannot blend into one another. The most
notable point is that loop quantum gravity as one fundamental theory
of quantum gravity is substantially testable with an observational
justification if primordial black holes are ever found.

This research was supported by Perimeter Institute for Theoretical
Physics.  Research at Perimeter Institute is supported by the
Government of Canada through Industry Canada and by the Province of
Ontario through the Ministry of Research and Innovation.


\begin{thebibliography}{}

\bibitem{Bojowald:2001xe}
  M.~Bojowald,
  Phys.\ Rev.\ Lett.\  {\bf 86}, 5227 (2001)
  [arXiv:gr-qc/0102069].



\bibitem{Ashtekar:1997yu}
  A.~Ashtekar, J.~Baez, A.~Corichi and K.~Krasnov,
  Phys.\ Rev.\ Lett.\  {\bf 80}, 904 (1998)
  [arXiv:gr-qc/9710007].




\bibitem{Ashtekar:surface}
 A.~Ashtekar and J.~Lewandowski,
  Class.\ Quant.\ Grav.\  {\bf 14}, A55 (1997).
  [arXiv:gr-qc/9602046];
S.~Frittelli, L.~Lehner and C.~Rovelli,
  Class.\ Quant.\ Grav.\  {\bf 13}, 2921 (1996)
  [arXiv:gr-qc/9608043];
  C.~Rovelli and L.~Smolin,
  Nucl.\ Phys.\ B {\bf 442}, 593 (1995)
  [Erratum-ibid.\ B {\bf 456}, 753 (1995)].
  [arXiv:gr-qc/9411005].



\bibitem{Ansari:2006vg}
  M.~H.~Ansari,
  Nucl.\ Phys.\  B {\bf 783}, 179 (2007)
  [arXiv:hep-th/0607081].


\bibitem{Rovelli}
  C.~Rovelli,
  Phys.\ Rev.\ Lett.\  {\bf 77}, 3288 (1996)
  [arXiv:gr-qc/9603063].


\bibitem{Ansari:2006cx}
  M.~H.~Ansari,
  arXiv:gr-qc/0603121.

\bibitem{Beckman:2001qs}
  D.~Beckman, D.~Gottesman, M.~A.~Nielsen and J.~Preskill,
  Phys.\ Rev.\  A {\bf 64}, 052309 (2001)
  [arXiv:quant-ph/0102043];
 T. Eggeling, D. Schlingemann, and R.~F. Werner,
[arXiv.org:quant-ph/0104027]; B. Schumacher, M. D. Westmoreland,
Quant. Info. Proc. 4, 13 (2005)
 [arXiv.org:quant-ph/0406223].


 \bibitem{sorkin}
 R. D. Sorkin, "Ten theses on black hole entropy," Stud.Hist. Philos. Mod. Phys. \textbf{36}, 291 (2005)

\bibitem{Bekenstein:1995ju}
  Bekenstein et. al.
  Phys.\ Lett.\ B {\bf 360}, 7 (1995) [arXiv:gr-qc/9505012].

\bibitem{hawking}
S. Hawking, Nature \textbf{248}, 30 (1974), S. W. Hawking, Commun.
Math. Phys. \textbf{43}, 199 (1975).


\bibitem{gerlach}
U. Gerlach, preceding paper, Phys. Rev. D \textbf{14}, 1479 (1976).



\end{thebibliography}
\end{document}